\documentclass[namedreferences]{SolarPhysics}
\usepackage[optionalrh]{spr-sola-addons} 
\usepackage{graphicx}        
\usepackage{color}           
\usepackage{url}             

\newcommand{\etal}{{\it et al.}}

\renewcommand{\vec}[1]{ {\mathbf #1} }

\newcommand{\grad}{ {\bf \nabla } }
\newcommand{\curl}{ {\bf \nabla} \times}

\newcommand{\avec}{ \vec A}
\newcommand{\nvec}{ \vec n}

\newcommand{\bvec}{ \vec B}

\newcommand{\aap}      {{\it Astron. Astrophys.}}

\newcommand{\apj}      {{\it Astrophys. J.}}
\newcommand{\apjl}{   {\it Astrophys. J. Lett.}}

\newcommand{\solphys}  {{\it Solar Phys.}}

\begin{document}

\begin{article}

\begin{opening}

\title{Algorithms of the Potential Field Calculation in a Three Dimensional Box}

\author{G. V.~\surname{Rudenko}$^{1}$\sep
               S. A.~\surname{Anfinogentov}$^{1}$\sep
       } \runningauthor{G.V.
Rudenko and S.A. Anfinogentov} \runningtitle{Super Fast and
Quality Azimuth Disambiguation}

   \institute{$^{1}$ Institute of Solar-Terrestrial Physics SB RAS, Lermontov St. 126,
Irkutsk 664033, Russia
                     email: \url{rud@iszf.irk.ru} email: \url{anfinogentov@iszf.irk.ru}\\}
\runningauthor{G.V. Rudenko and S.A. Anfinogentov}
\runningtitle{Extrapolation of a Potential Field in a 3D
Rectangular Box}

\begin{abstract}

Calculation of the potential field inside a three-dimensional box with the normal magnetic field component given on all boundaries is needed for estimation of important quantities related to the magnetic field such as free energy and relative helicity.
In this work we present an analysis of three methods for calculating potential field inside a three-dimensional box.
The accuracy and performance of the methods are tested on artificial models with a priori known solutions.

\end{abstract}
\end{opening}

\section{Introduction}
     \label{Section1}

Currently, many studies of the activity in solar active regions
rely on the photospheric magnetic field vector measurements and
the results of their subsequent processing. The latter includes
resolving the 180-degree ambiguity, non-linear force free field
(NLFFF) extrapolation, and estimating macroscopic magnetic
characteristics, such as free energy and relative helicity.
Research activities in  these areas are aimed at finding fast and
reliable methods for solving these problems on the basis of
various physical and mathematical concepts. During the last
decades, a significant progress has been made in these areas
\inlinecite{citeM1};\inlinecite{citeM2}; \inlinecite{citeM3};
\inlinecite{citeM4}; \inlinecite{citeM5}; \inlinecite{citeM6};
\inlinecite{citeM7}; \inlinecite{citeM8}; \inlinecite{citeM9};
\inlinecite{citeM10}; \inlinecite{citeM11}; \inlinecite{citeM12};
\inlinecite{citeMVal}; \inlinecite{citeM13}, providing  us with a
reliable (taking into account limitations of the available
measurements) description of the three-dimensional magnetic
structure and dynamics of real magnetic regions  in some cases
\inlinecite{citeA1}; \inlinecite{citeA2}; \inlinecite{citeA3};
\inlinecite{citeA4}; \inlinecite{citeA5}; \inlinecite{citeSun};
\inlinecite{citeKost}.

Despite the significant progress, the accuracy of existing
magnetic data processing methods requires improvement. For
instance, free energy estimations should have relative error of
less than  $10^{-2}$ to make tracking of flare energy release
possible \inlinecite{citeSun};\inlinecite{citeKost}. Similar
accuracy, presumably, is required for relative magnetic helicity
calculations. Current methods provide an accuracy of  $10^{-1}$ --
$10^{-2}$ in relative error \cite{citeSun}, which is not enough
for a reliable description of energy dynamics in real active
regions.

In this paper we present the analysis of three methods for potential field calculation in a 3D rectangular box.
This problem is one of key elements of the magnetic free energy and relative helicity estimation.
Seeking a precise computation of the potential field,  we, at least partially,  solve the problem of the accuracy of free energy estimation for solar active regions.
However, the potential field calculation is not the only source of uncertainty in this problem.
 A part is caused by the systematic errors in the measurements of the photospheric magnetic field and NLFFF extrapolation issues.
These errors are hard to estimate and, in principle, they can significantly exceed the required threshold.

We focus on two aspects of the considered numerical solutions: accuracy and performance.
We demonstrate that a balance  between the accuracy and the execution speed is achieved for one the methods, which can calculate  both the potential magnetic field and its vector potential.

\section{Derivation of the Potential Field, Scalar Potential, and Normalised Vector Potential for the Neuman BVP in a Rectangular Box }
      \label{Section2}

Let us consider the boundary value problem (BVP) in a rectangular box $\overline{V}$  with Neuman boundary conditions on its sides. In the following, we describe the algorithm for calculating three representation of this problem: scalar potential $\phi$, magnetic field
$\bvec$ and vector potential $\avec$. The problem consist of solving the following equations for $\phi$, $\bvec$ and $\avec$:

\begin{equation}\label{eq1}
 \ \Delta\phi=0, \ \  \ {\left(- \grad \phi \cdot
\nvec=g \right)}|_S,
\end{equation}
\begin{equation}\label{eq2}
 \left\{%
\begin{array}{ll}
     \ \curl{\bvec}={\vec 0} \\
     \ \grad \cdot {\bvec}=0 \\
     \ {\vec B}=-\grad\phi \\
\end{array}%
\right|_{\overline{V}}, \ \   \ {\left.  (\bvec \cdot \nvec=g)\right\vert}_S ,
\end{equation}

\begin{equation}\label{eq3}
 \left\{%
\begin{array}{ll}
     \  \Delta\avec={\bf 0} \\
     \ {\bf \nabla}\cdot\avec=0 \\
     \ \bvec={\bf \nabla}\times \avec \\
     \ {\bf \nabla}\times \avec=-{\bf \nabla}\phi \\
\end{array}%
\right|_{\overline{V}}, \ \
\begin{array}{ll} \ \ \
  \ {\left.  ({\bf \nabla}\times \avec \cdot \nvec=g)\right\vert}_S \\
\mathrm{Option} \   \ {\left.  (\avec \cdot \nvec=0)\right\vert}_S \\
\end{array}%
.
\end{equation}
Here $\nvec$ is the outward normal to the boundary of
$\overline{V}$, and $g$ is a normal field component given on the boundary surface  $S$:
$V\cup S=\overline{V}(x,y,z)$, $\left(x\in [0,L_x]; \ y\in
[0,L_y]; \ z\in [0,L_z]\right)$.  We assume that the divergence free condition for $g$ is satisfied:
\begin{equation}\label{eq4}
\int\limits_S g\mathrm{d}s=0.
\end{equation}
In the following, we consider the above BVP  (Equations (\ref{eq1}) -- (\ref{eq3})) without imposing the optional boundary condition in Equation (\ref{eq3}).
However this condition can be satisfied by adding to the calculated vector potential a harmonic function, found using the algorithm for solving the BVP formulated in Equation (\ref{eq2}).

Our algorithm is based on the reduction of the original problem to a series of BVPs with additional  to Equation (\ref{eq4}) the  requirement of a flux balance:
\begin{equation}\label{eq5}
\int\limits_S \widetilde{g}_i\mathrm{d}s=0 \ \ (i=1,2,...,6):  \ \left\{
\begin{array}{ll}
\widetilde{g}_i({\vec r})\neq 0; {\vec r} \in S_i\\
 \widetilde{g}_i({\vec r})= 0; {\vec r} \notin S_i
\end{array} \right. .
\end{equation}
Thus, the magnetic flux  must be balanced for each side of the rectangular volume $\overline{V}$ individually.
In the general case, the boundary conditions do not satisfy this requirement (Equation (\ref{eq5})), but we  always can represent  an arbitrary
boundary condition in the following form:
\begin{equation}\label{eq6}
g=g_a+\sum\limits_{i=1}^6\widetilde{g}_i,
\end{equation}
where  $g_a$ is the boundary condition  for a potential field
$\bvec_a$ that compensates the unbalanced flux $G_i$ on each side $i$:
\begin{equation}\label{eq7}
\int\limits_{S_i}g_a\mathrm{d}s=\int\limits_{S_i}g\mathrm{d}s\equiv G_i.
\end{equation}

Construction of such a compensating field $(\phi_a, \bvec_a, \avec_a)$ in analytical form is the first essential step of our algorithm.
When the compensating field is found, one can represent the general solution in the following form:

\begin{equation}\label{eq8}
\begin{array}{ll}
 \ \phi=\phi_a+\sum\limits_{i=1}^6\phi_i, \\
 \ \bvec=\bvec_a+\sum\limits_{i=1}^6\bvec_i, \\
 \ \avec=\avec_a+\sum\limits_{i=1}^6\avec_i,
\end{array}
\end{equation}
where $(\phi_i, \bvec_i, \avec_i)$ are the solutions of
the BVP$_i$ (Equations (\ref{eq1}) -- (\ref{eq4})) for six sides of the
volume $\overline{V}$.

\subsection{Construction of a Compensating Potential Field}
  \label{Section2.1}

Let us find the compensating potential field $(\phi_a, \bvec_a,
\avec_a)$  as a linear combination of five  harmonic solutions.
\begin{equation}\label{eq9}
\begin{array}{ll}
 \ \phi_a=\sum\limits_{j=1}^5m_j\phi_a^j, \\
 \ \bvec_a=\sum\limits_{j=1}^5m_j\bvec_a^j, \\
 \ \avec_a=\sum\limits_{j=1}^5m_j\avec_a^j.
\end{array}
\end{equation}

The essential requirement to the harmonic fields $\bvec_a^j$ is linear independence.
Hence, one can select an arbitrary set of linearly independent harmonic functions.
For the case of simplicity, we have selected them to be simple quadratic linearly independent polynomials:
\begin{equation}\label{eq10}
\begin{array}{ll}
\phi_a^1=-x, \\
\phi_a^2=-y, \\
\phi_a^3=-z, \\
\phi_a^4=-x^2+z^2, \\
\phi_a^5=-y^2+z^2;
\end{array}
\end{equation}
\begin{equation}\label{eq11}
\begin{array}{ll}
\bvec_a^1=(1,0,0), \\
\bvec_a^2=(0,1,0), \\
\bvec_a^3=(0,0,1), \\
\bvec_a^4=(2x,0,-2z), \\
\bvec_a^5=(0,2y,-2z);
\end{array}
\end{equation}
\begin{equation}\label{eq12}
\begin{array}{ll}
\avec_a^1=(0,0,y), \\
\avec_a^2=(z,0,0), \\
\avec_a^3=(0,x,0), \\
\avec_a^4=(0,-2xz,0), \\
\avec_a^5=(2yz,0,0).
\end{array}
\end{equation}
Since the selection of Equation (\ref{eq10}) is arbitrary, one can freely select another set.
For example, $-x^2+z^2$  and $-y^2 + z^2$ can be replaced by $xz$  and $yz$.

The fluxes through the box sides  can be calculated analytically for each basis solution:
\begin{equation}\label{eq13}
\int\limits_{S_i}g_a^j\mathrm{d}s=\int\limits_{S_i}\bvec_a^j\cdot\nvec \mathrm{d}s=G_{ai}^j.
\end{equation}

For instance, let us calculate the flux of the basis field $\bvec_a^3$ through  the bottom side of the box $(x, y,0)$.
Applying Equation (\ref{eq13}), we get
$$G_{a1}^3=\int\limits_{0}^{L_x}\int\limits_{0}^{L_y}B_{az}^3\times(-1)\mathrm{d}x\mathrm{d}y=-\int\limits_{0}^{L_x}\int\limits_{0}^{L_y}\mathrm{d}x\mathrm{d}y=L_xL_y.$$

After calculation of the $G_{ai}^j$ values,  we find the unknown coefficients $m_j$  by solving a linear system of algebraic equations, representing the flux balance conditions on five sides (Equation (\ref{eq7})):
\begin{equation}\label{eq14}
G_{ai}^jm_j=G_i \ \ \ (i=1,...,5; j=1,...,5).
\end{equation}
The condition in Equation (\ref{eq14}) in combination with the divergence-free condition (Equation (\ref{eq4})) automatically provides flux balance (Equation (\ref{eq7})) on all
six sides.

The found solution $(\phi_a, \bvec_a, \avec_a)$
allows us to obtain modified boundary conditions $\widetilde{g}_i$ balanced on each side of the volume:
\begin{equation}\label{eq15}
\widetilde{g}_i=g_i-g_{ai} \ \ \ (i=1,...,6).
\end{equation}
\subsection{Solution of a BVP with Non-zero Boundary Conditions on One Side of a Rectangular Box}
We present two different approaches for solving this problem: an analytic and a numerical one.
In the first case, we derive exact analytic  equations for the magnetic field components, and its scalar and vector potentials.
In practical implementations the integrals which appear are computed directly with the use of 2D local quadratic splines allowing for analytical integration of their convolution with trigonometric functions.
The second approach solves the problem in the discrete Fourier space and uses the fast Fourier transform (FFT) to speed up the computation.

\subsubsection{Analytical Solution}
  \label{Section2.2}
Let us consider in detail the solution of one of the sub-problems
$($BVP$)_i$ $(g = \widetilde{g}_i)$ with non-zero boundary conditions $\widetilde{g}_i$  given on the bottom
side of the computational box $(x, y, 0)$.
For other sides, the solution can be found in the same way.
We search for the solution in the form of a   decomposition into a set of harmonic basis functions  of the following form:
\begin{equation}\label{eq16}
\phi=\sum\limits_{m=0}^\infty\sum\limits_{n=0}^\infty
p_{mn}\phi_{mn}=\sum\limits_{m=0}^\infty\sum\limits_{n=0}^\infty
p_{mn}C_x^mC_y^nZ^{mn},
\end{equation}
\begin{equation}\label{eq17}
\begin{array}{ll}
B_x=\sum\limits_{m=0}^\infty\sum\limits_{n=0}^\infty
p_{mn}(B_x)_{mn}=-\sum\limits_{m=0}^\infty\sum\limits_{n=0}^\infty
p_{mn}C_x^{'m}C_y^nZ^{mn}, \\
B_y=\sum\limits_{m=0}^\infty\sum\limits_{n=0}^\infty
p_{mn}(B_y)_{mn}=-\sum\limits_{m=0}^\infty\sum\limits_{n=0}^\infty
p_{mn}C_x^mC_y^{'n}Z^{mn}, \\
B_z=\sum\limits_{m=0}^\infty\sum\limits_{n=0}^\infty
p_{mn}(B_z)_{mn}=-\sum\limits_{m=0}^\infty\sum\limits_{n=0}^\infty
p_{mn}C_x^mC_y^nZ^{'mn};
\end{array}
\end{equation}
\begin{equation}\label{eq18}
\begin{array}{ll}
A_x=\sum\limits_{m=0}^\infty\sum\limits_{n=0}^\infty
p_{mn}(A_x)_{mn}=-\sum\limits_{m=0}^\infty\sum\limits_{n=0}^\infty
p_{mn}C_x^mC_y^{'n}Z^{'mn}\frac{\pi n}{q^2L_y}, \\
A_y=\sum\limits_{m=0}^\infty\sum\limits_{n=0}^\infty
p_{mn}(A_y)_{mn}=\sum\limits_{m=0}^\infty\sum\limits_{n=0}^\infty
p_{mn}C_x^{'m}C_y^nZ^{'mn}\frac{\pi m}{q^2L_x}, \\
A_z=(A_z)_{mn}=0,
\end{array}
\end{equation}
where
\begin{equation}\label{eq19}
q_{mn}=\sqrt{\left(\frac{\pi m}{L_x}\right)^2+\left(\frac{\pi
n}{L_y}\right)^2},
\end{equation}
\begin{equation}\label{eq20}
\begin{array}{ll}
C_x^m=\cos \left( \frac{\pi m}{L_x}x\right), \ \
C_x^{'m}=-\frac{\pi m}{L_x}\sin \left( \frac{\pi m}{L_x}x\right),
\\
C_y^n=\cos \left( \frac{\pi n}{L_y}y\right), \ \
C_y^{'n}=-\frac{\pi n}{L_y}\sin \left( \frac{\pi n}{L_y}y\right),
\\
Z^{mn}=\left(1-\frac{1}{2}\delta_{0m}-\frac{1}{2}\delta_{0n}\right)\frac{4}{L_xL_yq_{mn}}e^{-q_{mn}z}\frac{1+e^{-2q_{mn}(L_z-z)}}{1-e^{-2q_{mn}L_z}},
\\
Z^{'mn}=-\left(1-\frac{1}{2}\delta_{0m}-\frac{1}{2}\delta_{0n}\right)\frac{4}{L_xL_y}e^{-q_{mn}z}\frac{1-e^{-2q_{mn}(L_z-z)}}{1-e^{-2q_{mn}L_z}}.
\end{array}
\end{equation}
Each partial harmonic solution $(\phi_{mn}, \bvec_{mn}, {\bf
A}_{mn})$ satisfies Equations (\ref{eq1}) -- (\ref{eq4}) with $g \neq 0$ on the
side $(x, y, 0)$ and $g = 0$ on the other sides. The
coefficients $p_{mn}$ are determined by the following formula:
\begin{equation}\label{eq21}
p_{mn}=-\int\limits_0^{L_x}\int\limits_0^{L_y}(B_z)_{mn}\widetilde{g}_i\mathrm{d}x\mathrm{d}y,
\ \ (z=0).
\end{equation}
The solutions
$(\widetilde{\phi}_i, \widetilde{\bvec}_i, \widetilde{\avec}_i)$  found in this way for each side $i$ allow us to obtain the final solution of the problem (Equations (\ref{eq1}) -- (\ref{eq4}))
as the following sum:
\begin{equation}\label{eq22}
(\phi, \bvec, \avec)=(\phi_a, \bvec_a, \avec_a)+\sum\limits_{i=1}^6(\widetilde{\phi}, \widetilde{\bvec},
\widetilde{\avec}).
\end{equation}

\subsection{Variants of the BVP Numerical Implementation }
  \label{Section2.3}
  In a practical implementation of this scheme, the number of terms in the expansion Equations (\ref{eq16}) -- (\ref{eq18}) can be  limited by the size of the grid $(x,y)$.
  In the following, we benchmark two different implementations of our method.
  In the first implementation (CASE I), the coefficients $p_{mn}$  are  calculated  by the direct integration of Equation  (\ref{eq21}).
  To improve the accuracy, we present the boundary condition $\widetilde{g}_i$ in the form of 2D  local quadratic splines allowing for analytical integration of their convolution with trigonometric functions.
  In our second implementation, we calculate $p_{mn}$ coefficients with the use of the FFT (CASE~II).
Then the components of the  field and vector potential are computed in terms of FFT coefficients in a form similar to Equations (\ref{eq17}) and (\ref{eq18}).

As a reference, we also include in our test the results of a proprietary Poisson solver built into the Intel MKL library (CASE~III).
Since the above package allows us to calculate only the scalar potential $\phi$, the  magnetic field $\bvec$ is computed as the gradient of ${\phi}$ (Equation (\ref{eq2}), the third equation of the system) using the finite difference scheme.  The vector potential $\avec$ is calculated from the obtained field $\bvec$ with the use of the algorithm developed by \cite{citeMVal}. Note that  the  vector potential $\avec$ computed by this algorithm generally  does not satisfy the conditions  Equation (\ref{eq3}) (the second equation of the system)  and Equation (\ref{eq3}) (the first equation of the system).

\section{Numerical Tests}
\label{Section3}

We test our method on two models: the field of two magnetic charges situated outside the box and the magnetic field of a typical solar active region obtained by potential field extrapolation from the normal component  measured at the photospheric level under the condition of a finite field at  infinite distance.

\subsection{Notations and Metrics}
\label{Section3.1}
Let us introduce  the following notations and metrics:
\begin{itemize}
    \item Numerical solution: $(\phi_{\rm calc}, \bvec_{\rm calc}, \avec_{\rm calc})$
    \item Model solution: $(\phi_{\rm model}, \bvec_{\rm model}, \avec_{\rm model})$
    \item Local relative error:
        \begin{equation}\label{eq23}
        \epsilon=\frac{\left|\bvec_{\rm  calc}-\bvec_{\rm model}\right|}{\left|\bvec_{\rm model}\right|}.
        \end{equation}
    \item Average relative error:
        \begin{equation}\label{eq24}
        \left<\epsilon\right>_{\rm mean}=\mathrm{mean}(\epsilon).
        \end{equation}
    \item Median relative error:
        \begin{equation}\label{eq25}
        \left<\epsilon\right>_{\rm median}=\mathrm{median}(\epsilon).
        \end{equation}
    \item Weighted mean relative error:
        \begin{equation}\label{eq26}
        \left<\epsilon\right>_{\rm  w}=\frac{\mathrm{mean}(\left|\bvec_{\rm  model}\right|\epsilon)}{\mathrm{mean}(\left|\bvec_{\rm model}\right|)}.
        \end{equation}
    \item The average error of the calculated scalar potential:
        \begin{equation}\label{eq27}
        \left<\epsilon\right>_{\rm w}=\mathrm{mean}\left(2\frac{\left|\phi_{\rm calc}-\phi_{\rm model}\right|}{\left|\phi_{\rm calc}\right|+\left|\phi_{\mathrm{model}}\right|}\right).
        \end{equation}
    \item Average relative error:
        \begin{equation}\label{eq28}
        \left<\epsilon_z\right>_{\rm mean}=\mathrm{mean}(\epsilon): \ (z=\mathrm{const}).
        \end{equation}
    \item Median relative error:
        \begin{equation}\label{eq29}
        \left<\epsilon_z\right>_{\rm median}=\mathrm{median}(\epsilon): \ (z=\mathrm{const}).
        \end{equation}
    \item Weighted mean relative error:
        \begin{equation}\label{eq30}
        \left<\epsilon_z\right>_{\rm  w}=\frac{\mathrm{mean}(\left|\bvec_{\rm  model}\right|\epsilon)}{\mathrm{mean}(\left|\bvec_{\rm model}\right|)}: \
        (z=\mathrm{const}).
        \end{equation}
    \item Relative amount of nodes with an error is less than the $\epsilon$:
        \begin{equation}\label{eq31}
        f(\epsilon)=\frac{n(<\epsilon)}{n_{\rm total}}.
        \end{equation}
    \item Magnetic field energy inside the computational box $V$ (without boundaries):
        \begin{equation}\label{eq32}
        E=\int\limits_V\frac{\left|\bvec\right|^2}{8\pi}\mathrm{d}v.
        \end{equation}
    \item Energy, calculated by the virial theorem. Here $S'$ is the boundary of the inner region $V'$ and $\nvec$ is the outward normal to the boundary $S'$:
        \begin{equation}\label{eq33}
        E_{\rm virial}=\frac{1}{4\pi}\int\limits_{S'}\left(\frac{1}{2}\left|\bvec \right|^2(\nvec\cdot{\vec r})-(\nvec\cdot\bvec)({\bf r}\cdot\bvec)\right)\mathrm{d}s.
        \end{equation}
    \item Energy, calculated from the normal  component of the magnetic field and scalar potential given on the Surface $S'$:
        \begin{equation}\label{eq34}
        E_{\phi}=\frac{1}{8\pi}\int\limits_{S'}\phi(\nvec\cdot\bvec)\mathrm{d}s.
        \end{equation}

\end{itemize}

\subsection{Model of Two Charges}
\label{Section3.2} The model of two charges ($M_{\rm charges}$) is defined analytically as follows:

\begin{equation}
    \phi_{\rm model}=-\frac{q_1}{|{\vec r}-{\vec r}_1|}-\frac{q_2}{|{\vec r}-{\vec r}_2|},
\end{equation} \label{eq35_start}
\begin{equation}
    \bvec_{\rm  model}=\frac{q_1({\vec r}-{\vec r}_1)}{|{\vec r}-{\vec r}_1|^3}+\frac{q_2({\vec r}-{\vec r}_2)}{|{\vec r}-{\vec r}_2|^3},
\end{equation}
whith
$q_1=1000$, $q_2=-1500$,
${\vec r}_1=(-0.2,-0.18,-0.22)$, ${\vec r}_2=(0.2,0.18,-0.22)$,
$\overline{V}=[-1,0.233]\times [-1,0.953] \times [0,1.610]$,
and grid size
$
\left[ n_x, n_y, n_z \right]=[37,58,48].
$

\begin{figure}
\centerline{\includegraphics[width=1.\textwidth,clip=]{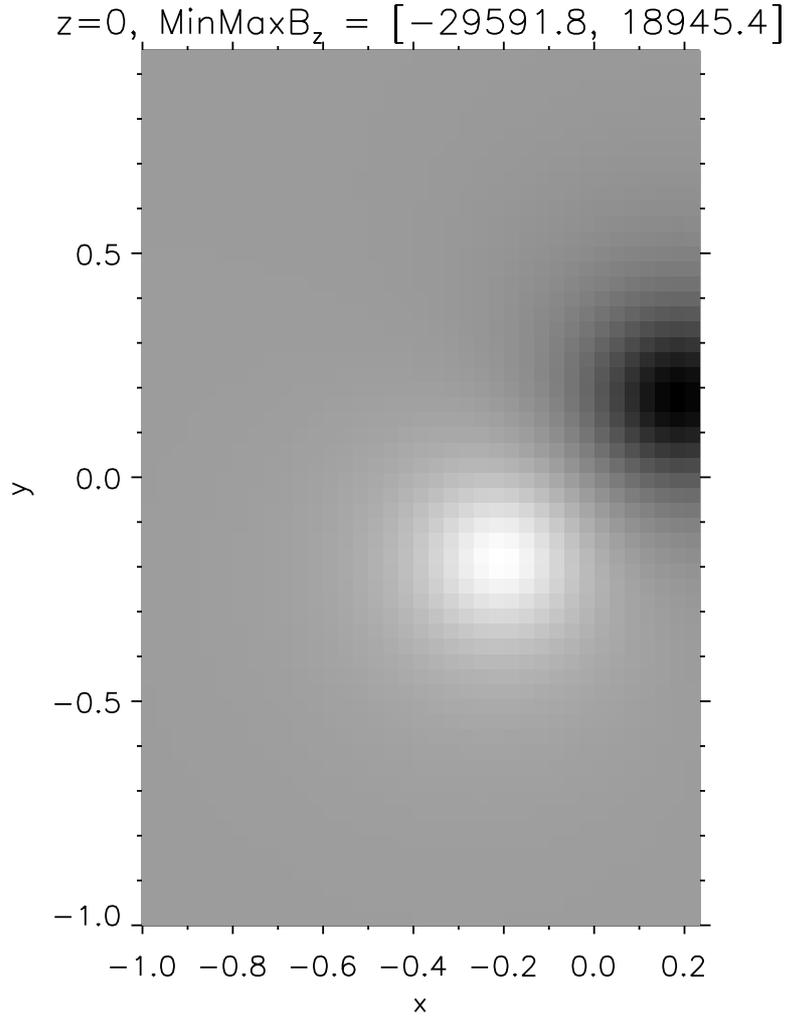}}
\caption{Normal component of the magnetic field at the lower boundary of the $M_{\rm charges}$ model.} \label{fig1}
\end{figure}

The model parameters are selected to give a magnetic field configuration with large values on one of the side faces of the box, as demonstrated in Figure \ref{fig1}.
\begin{figure}
\centerline{\includegraphics[width=1.\textwidth,clip=]{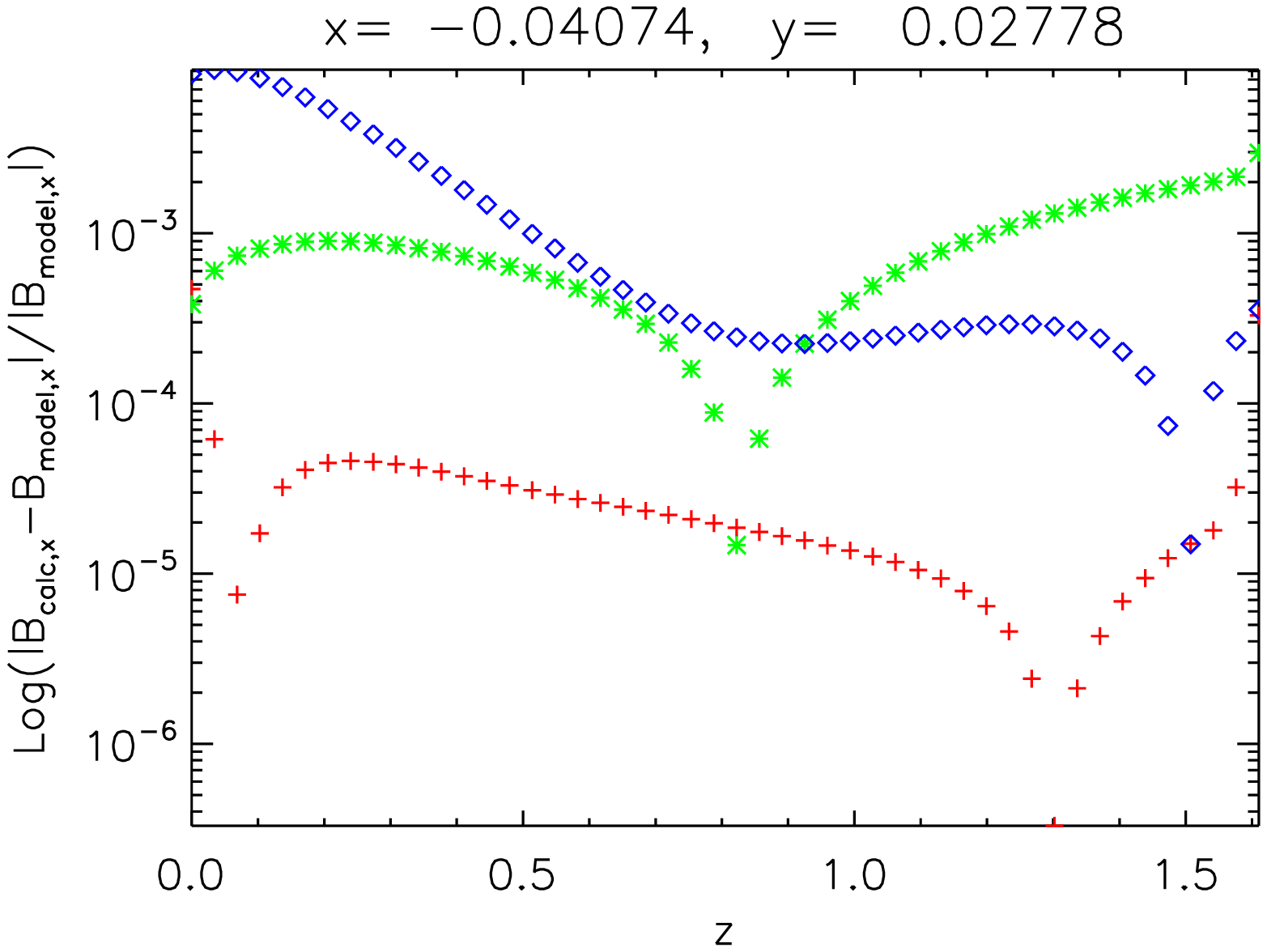}}
\caption{Relative errors in the $B_x$  component extracted from a  one-dimensional cross-section ($x  = -0.04074$, $y = 0.02778$) of the computational box for the solutions obtained with the Case I (red crosses), Case II (green asterisks) and case III (blue diamonds)  methods.} \label{fig2}
\end{figure}
\begin{figure}
\centerline{\includegraphics[width=1.\textwidth,clip=]{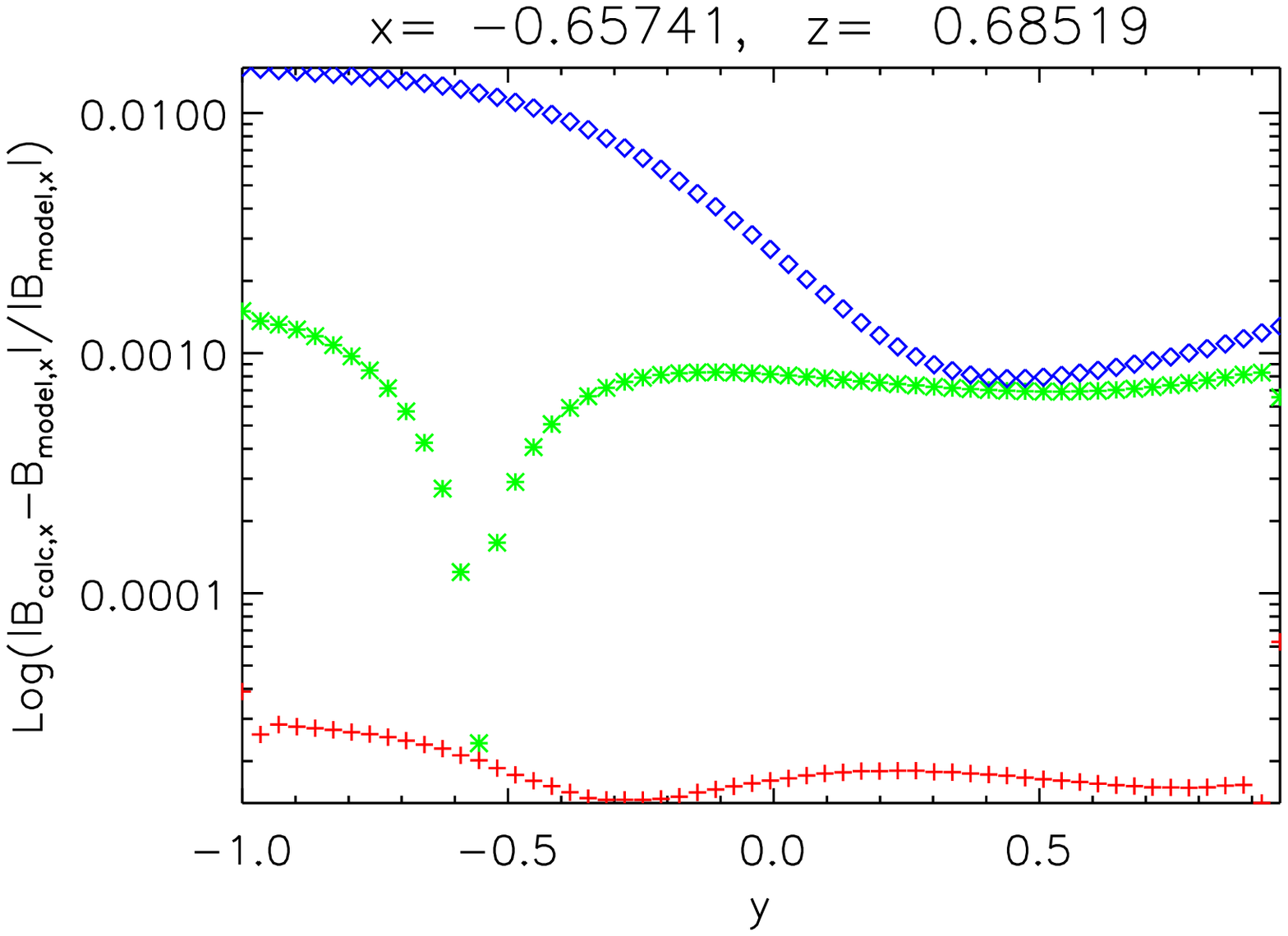}}
\caption{Relative errors in the $B_x$  component extracted from  a  one-dimensional cross-section ($x  = -0.65741$, $z = 0.68519$) of the computational box for the solutions obtained with the Case I (red crosses), Case II (green asterisks) and case III (blue diamonds)  methods.} \label{fig3}
\end{figure}
\begin{figure}
\centerline{\includegraphics[width=1.\textwidth,clip=]{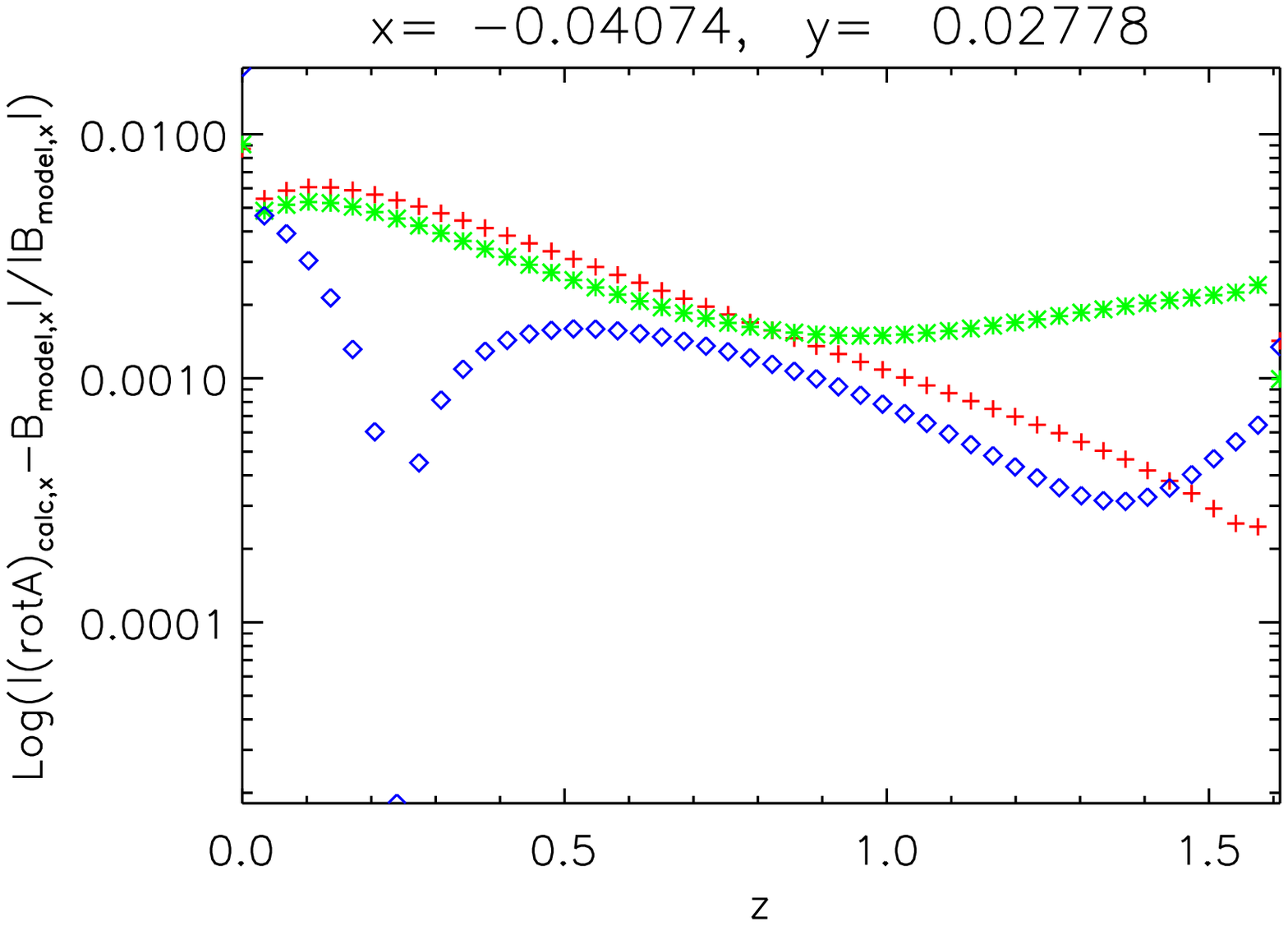}}
\caption{Relative errors in the $B_x$  component reconstructed from the vector potential and extracted from a  one-dimensional cross-section ($x  = -0.04074$, $y = 0.02778$) of the computational box for the solutions obtained with the Case I (red \textbf{crosses}), Case II (green asterisks) and Case III (blue diamonds)  methods.} \label{fig4}
\end{figure}
\begin{figure}
\centerline{\includegraphics[width=1.\textwidth,clip=]{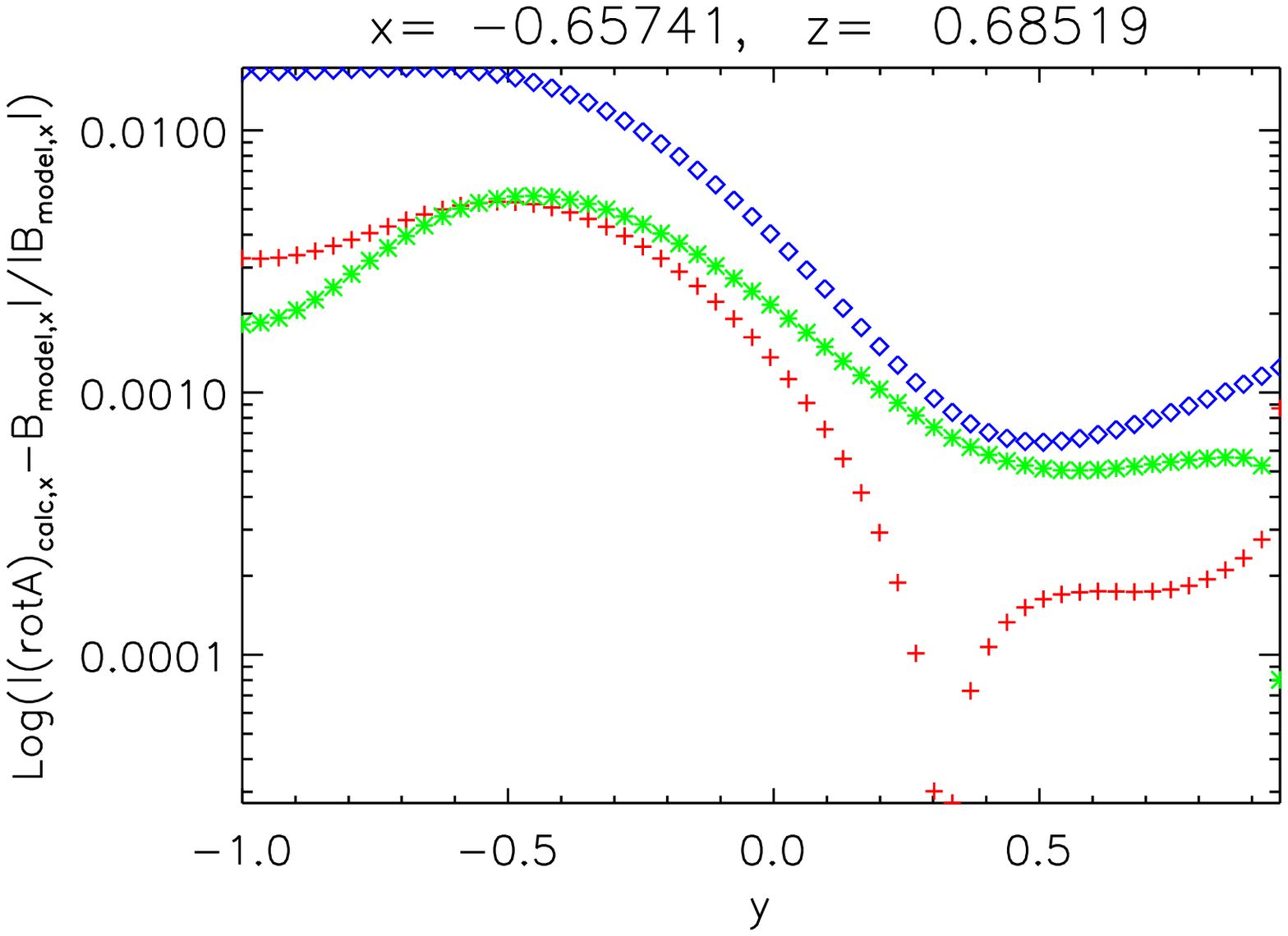}}
\caption{Relative errors in the $B_x$  component reconstructed from the vector potential and extracted from a  one-dimensional cross-section ($x  = -0.65741$, $z = 0.68519$) of the computational box for the solutions obtained with the Case I (red \textbf{crosses}), Case II (green asterisks) and case III (blue diamonds)  methods.} \label{fig5}
\end{figure}

For the first assessment of the method's accuracy, we present the relative errors in the $B_x$  component extracted from two arbitrarily  chosen  one-dimensional cross-sections of the computational box (See Figures \ref{fig2} -- \ref{fig5}).
All  methods show good agreement with the model, both inside the box, and on its boundaries.
The best accuracy is demonstrated by Case I, followed by the Case II.
As expected, the accuracy of the field calculated from the vector potential $({\bf \nabla} \times \avec)_{{\rm calc}, x}$ using the finite difference method is  similar for all 3 cases.

The global metrics Equations (\ref{eq24}) -- (\ref{eq27}) characterizing the
accuracy of the three  BVP solutions in the volume $V$ are
presented in Tables \ref{table1} and \ref{table2}. The metrics for the boundary $S$ are given in
Tables \ref{table3} and \ref{table4}.
\begin{table}
\caption{Model $M_\mathrm{charges}$: volume metrics of $\bvec_{\rm  calc}$
and the scalar potential $\phi$ computed for the entire volume $V$.}
 \label{table1}
\begin{tabular}{cccccc}
\hline
  & $\left<\epsilon\right>_{\rm mean}$ & $\left<\epsilon\right>_{\rm median}$& $\epsilon_{\rm max}$& $\left<\epsilon\right>_{\rm w}$ & $\left<\epsilon\right>_{\phi}$\\
 \hline
Case I   &$9.734\times 10^{-5}$&$7.850\times 10^{-5}$&$8.244\times 10^{-4}$&$9.734\times 10^{-5}$&$9.745\times 10^{-5}$  \\
Case II   &$1.111\times 10^{-3}$&$1.091\times 10^{-3}$&$6.401\times 10^{-3}$&$1.038\times 10^{-3}$&$2.925\times 10^{-3}$  \\
Case III   &$8.197\times 10^{-3}$&$6.935\times 10^{-3}$&$5.524\times 10^{-2}$&$5.343\times 10^{-3}$&$1.084\times 10^{-2}$  \\
 \hline
\end{tabular}
\end{table}

\begin{table}
\caption{Model $M_{\rm charges}$: surface metrics for $\bvec_{\rm  calc}$
and the scalar potential $\phi$ computed for the whole boundary $S$.}
 \label{table2}
\begin{tabular}{cccccc}
\hline
  & $\left<\epsilon\right>_{\rm mean}$ & $\left<\epsilon\right>_{\rm median}$& $\epsilon_{\rm max}$& $\left<\epsilon\right>_{\rm w}$ & $\left<\epsilon\right>_{\phi}$\\
\hline
Case I   &$1.043\times 10^{-3}$&$4.679\times 10^{-4}$&$2.594\times 10^{-2}$&$1.278\times 10^{-3}$&$1.419\times 10^{-4}$  \\
Case II   &$1.999\times 10^{-3}$&$1.487\times 10^{-3}$&$1.918\times 10^{-2}$&$2.331\times 10^{-3}$&$4.184\times 10^{-3}$  \\
Case III   &$1.293\times 10^{-2}$&$9.985\times 10^{-3}$&$1.847\times 10^{-1}$&$1.311\times 10^{-2}$&$1.445\times 10^{-2}$  \\
 \hline
\end{tabular}
\end{table}

\begin{table}
\caption{Model $M_{\rm charges}$: volume metrics  for $\curl\avec_{\rm calc}$ and the scalar potential $\phi$ computed for the entire volume $V$.}
 \label{table3}
\begin{tabular}{ccccc}
\hline
  & $\left<\epsilon\right>_{\rm mean}$ & $\left<\epsilon\right>_{\rm median}$& $\epsilon_{\rm max}$& $\left<\epsilon\right>_{\rm w}$ \\
 \hline
Case I   &$1.528\times 10^{-3}$&$1.050\times 10^{-3}$&$1.680\times 10^{-2}$&$3.401\times 10^{-3}$  \\
Case II   &$2.001\times 10^{-3}$&$1.702\times 10^{-3}$&$2.244\times 10^{-2}$&$3.558\times 10^{-3}$  \\
Case III   &$8.460\times 10^{-3}$&$6.525\times 10^{-3}$&$8.298\times 10^{-2}$&$5.726\times 10^{-3}$  \\
 \hline
\end{tabular}
\end{table}

\begin{table}
\caption{Model $M_{\rm charges}$: surface metrics  for $\curl\avec_{\rm calc}$ and the scalar potential $\phi$ computed for the boundary surface  $S$.}
 \label{table4}
\begin{tabular}{ccccc}
\hline
  & $\left<\epsilon\right>_{\rm mean}$ & $\left<\epsilon\right>_{\rm median}$& $\epsilon_{\rm max}$& $\left<\epsilon\right>_{\rm w}$ \\
 \hline
Case I   &$2.616\times 10^{-3}$&$1.600\times 10^{-3}$&$2.822\times 10^{-2}$&$8.622\times 10^{-3}$  \\
Case II   &$3.458\times 10^{-3}$&$2.322\times 10^{-3}$&$4.253\times 10^{-2}$&$8.909\times 10^{-3}$  \\
Case III   &$1.875\times 10^{-2}$&$1.426\times 10^{-2}$&$1.756\times 10^{-1}$&$1.932\times 10^{-2}$  \\
 \hline
\end{tabular}
\end{table}

Tables \ref{table1} and \ref{table2} confirm that  Case I provides the most accurate solution. Both volume and surface metrics are an order of magnitude better than in Case~II.
The Case~II method is more accurate than Case~III by an order of magnitude for most of the  metrics.
The metrics for the magnetic field reconstructed from the vector potential ${\bf \nabla}\times\avec_{\rm calc}$ (Tables
\ref{table3} and \ref{table4}) are  less precise due to the usage of the finite difference approximation. However, the relative errors of all methods are small  both in the volume and on the boundary, demonstrating sufficiently high quality for practical application.

\begin{figure}
\centerline{\includegraphics[width=1.\textwidth,clip=]{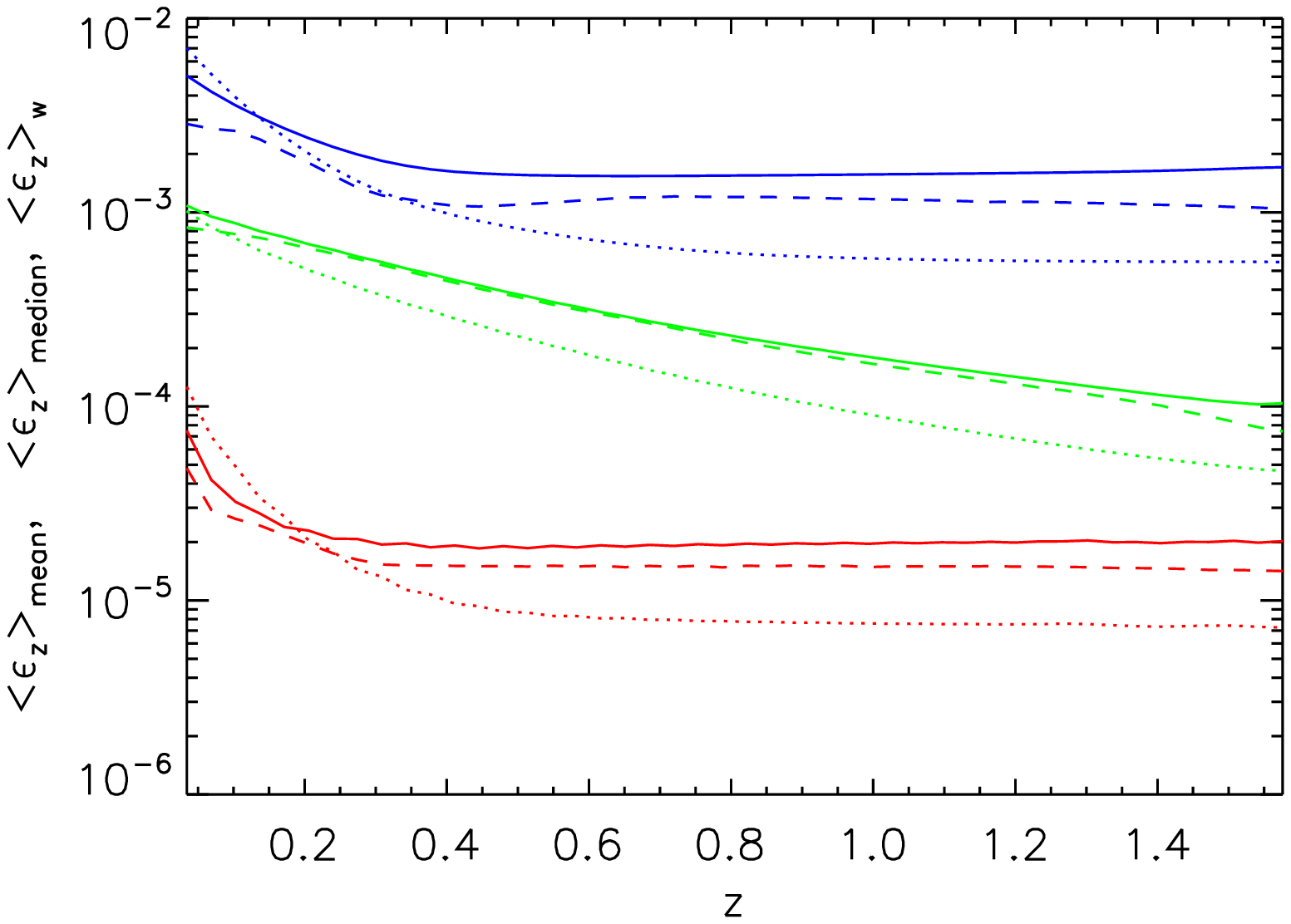}}
\caption{
Dependence of the  average (solid line), median (dashed line) and the field-weighted (dotted line) relative error with respect to the model field ($\bvec_{\rm model}$)  on the height ($z$) for the  potential magnetic field of the $M_{\rm charges}$ model, calculated with the Case I (red), Case II (green) and Case III (blue) methods from the Neumann boundary conditions.}  \label{fig6}
\end{figure}
\begin{figure}
\centerline{\includegraphics[width=1.\textwidth,clip=]{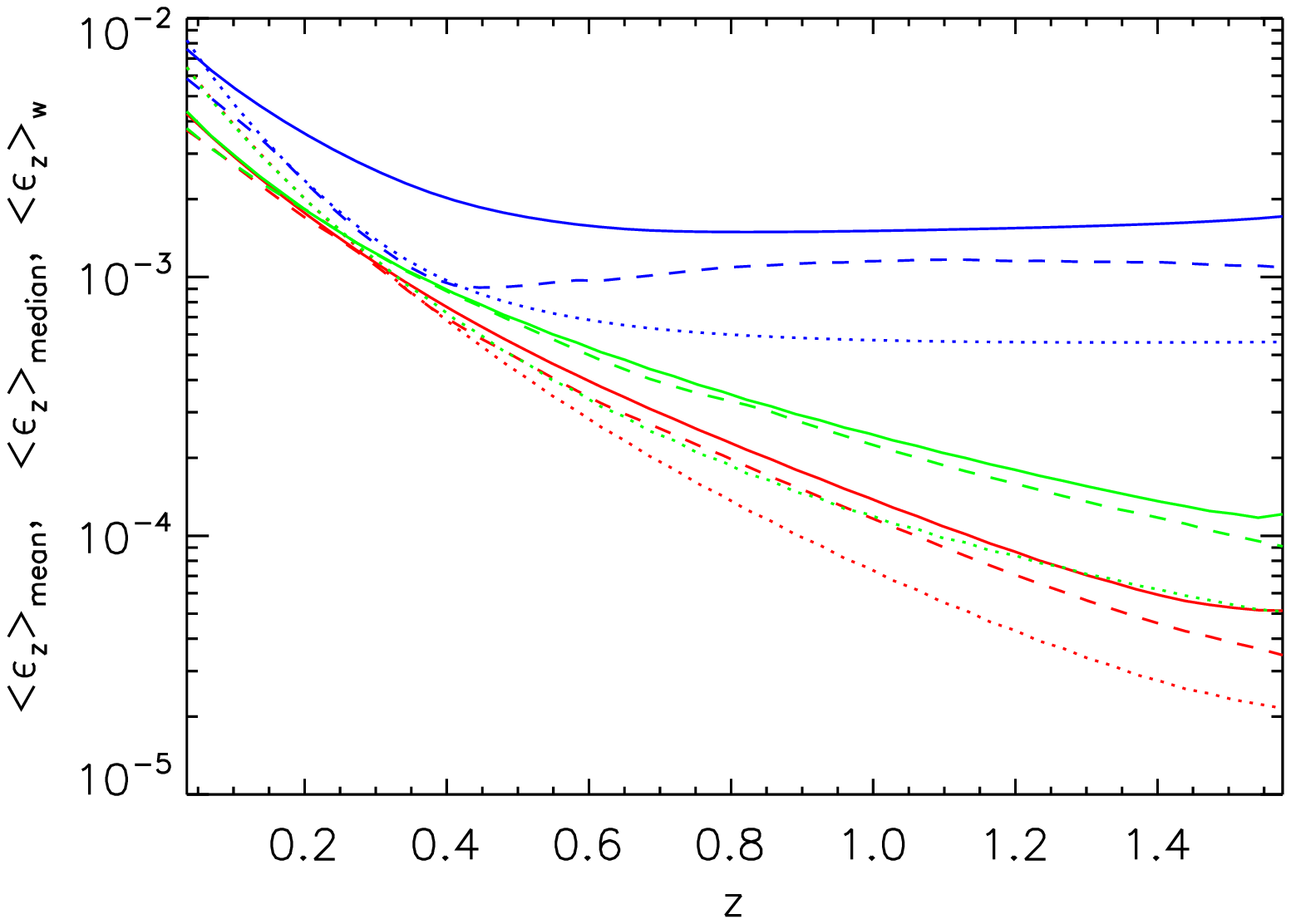}}
\caption{
    Dependence of the  average (solid line), median (dashed line) and the field-weighted (dotted line) relative error with respect to the model field ($\bvec_{\rm model}$)  on the height ($z$) for the  potential magnetic field of the $M_{\rm charges}$ model,  recovered from the vector potential (${\bf \nabla}\times\avec_{\rm calc}$) calculated with the Case I (red), Case II (green) and Case III (blue) methods from the Neumann boundary conditions.
} \label{fig7}
\end{figure}
\begin{figure}
\centerline{\includegraphics[width=1.\textwidth,clip=]{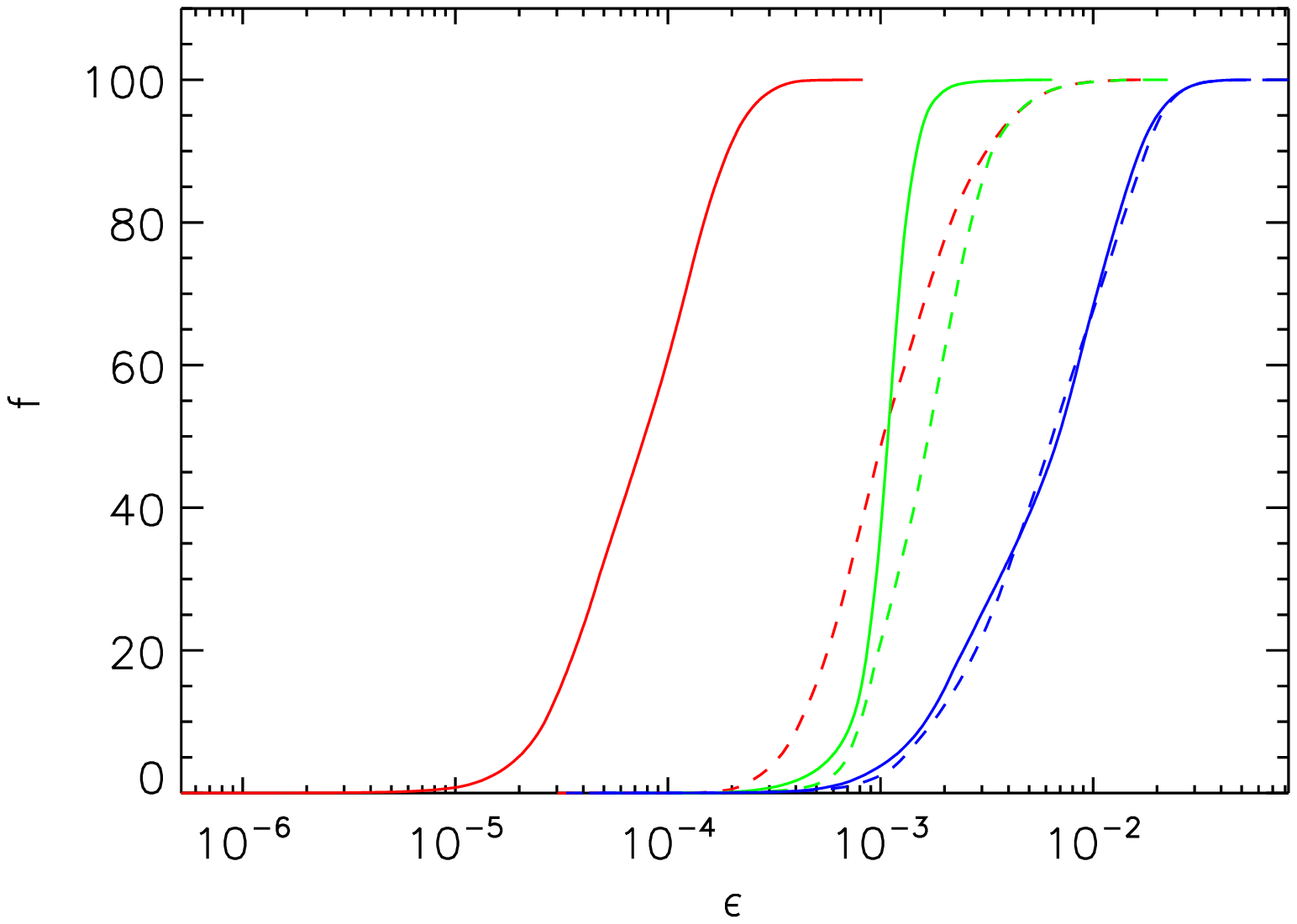}}
\caption{
    The distribution of the relative errors of the field $\bvec_{\rm calc}$ (solid line) and the field computed from the vector potential ${\bf \nabla}\times\avec_{\rm calc}$  (dashed line) with respect to the model field$\bvec_{\rm model}$ for the potential field of the $M_{\rm charges}$ model calculated with the Case I (red), Case II (green) and Case III (blue) methods from the Neumann boundary conditions.}
\label{fig8}
\end{figure}

The additional metrics  Equations (\ref{eq28}) --
(\ref{eq31}) shown in Figures \ref{fig6} -- \ref{fig8} demonstrate that the solution accuracy increases with height ($z$). Hence, all algorithms tend to produce more accurate solutions in areas of smooth and low magnitude fields.
Figure \ref{fig8} shows that the fraction of grid points with large relative errors ($> 1\%$) is extremely low in the entire volume  $V$.

\begin{table}
\caption{Model $M_{\rm charges}$: magnetic energy calculation results.}
 \label{table5}
\begin{tabular}{cccc}
\hline
  & $E$ & $E_\phi$& $E_{virial}$ \\
 \hline
Model   &$2.9780\times 10^{5}$&$2.9710\times 10^{5}$&$3.0072\times 10^{5}$  \\
Case I   &$2.9776\times 10^{5}$&$2.9705\times 10^{5}$&$3.0069\times 10^{5}$  \\
Case II   &$2.9727\times 10^{5}$&$2.9654\times 10^{5}$&$3.0020\times 10^{5}$  \\
Case III   &$2.9684\times 10^{5}$&$2.9673\times 10^{5}$&$3.0334\times 10^{5}$  \\
 \hline
\end{tabular}
\end{table}

In Table \ref{table5}, we compare the values of the total magnetic energy calculated in different ways: by integration over the volume $V$ (Equation (\ref{eq32})) and over the boundary $S$ (Equations (\ref{eq33}) and (\ref{eq34})).
Also, Table 5 gives an idea of how accurate are  the different algorithms for the calculation of the magnetic field energy that is  an important macroscopic parameter of  a solar active region.

\subsection{Model of a Real Active Region}
\label{Section3.3}

\begin{figure}
\centerline{\includegraphics[width=1.\textwidth,clip=]{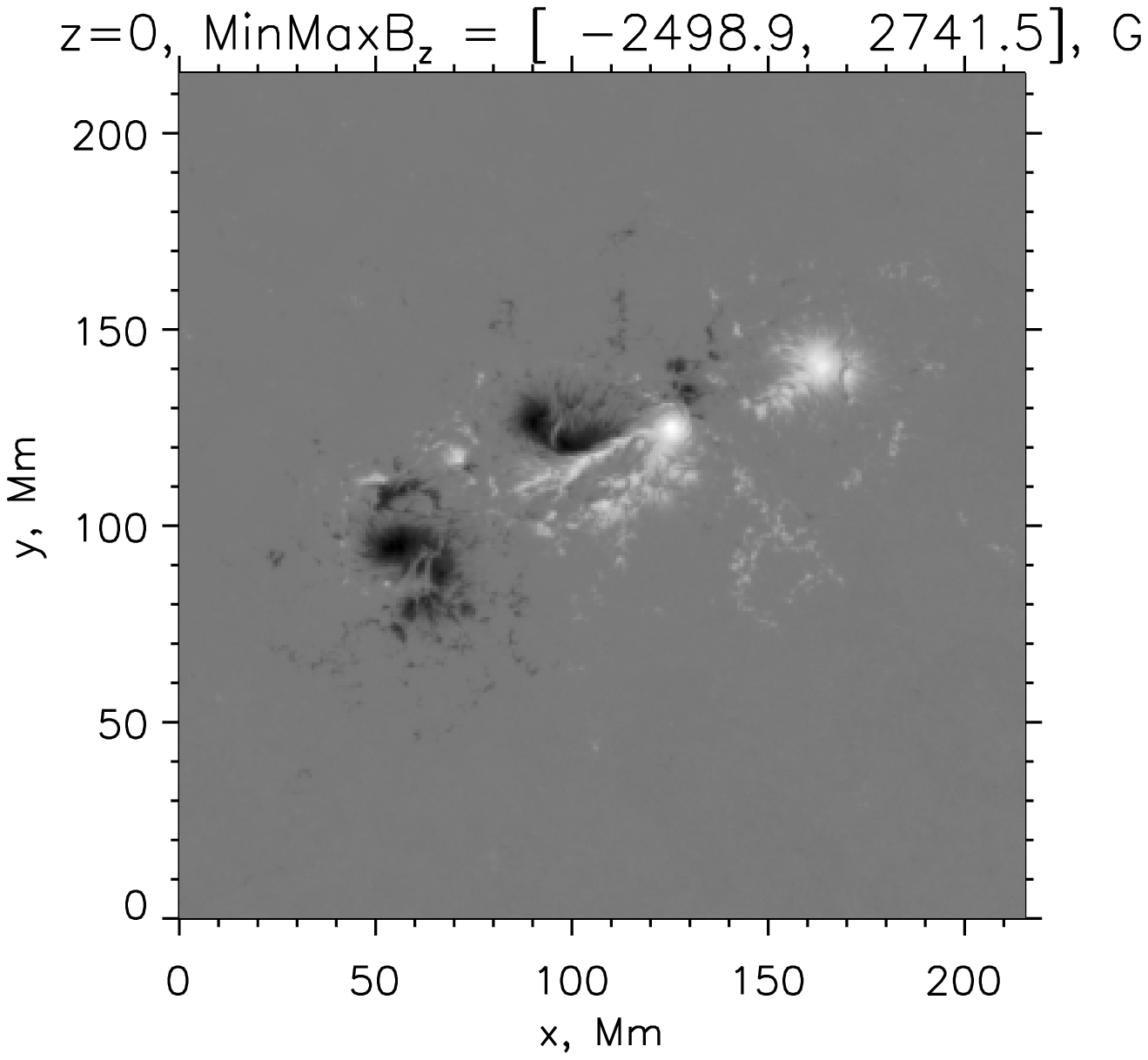}}
\caption{ Magnetic field normal component at the lower boundary of the
    $M_{\rm AR}$ model based on the  SDO/HMI photospheric magnetogram of AR 11158 observed at 2011-02-15 01:36:00 UT.
} \label{fig9}
\end{figure}

The realistic $M_{\rm AR}$ is built by the potential extrapolation of the magnetic field of AR 11158 from the normal magnetic field component observed at the photospheric  level by SDO/HMI at 2011-02-15 01:36:00 UT.
The potential field calculation is performed using the FFT based algorithm proposed by
\cite{1981A&A...100..197A}.
The size of the computational box is 300$\times$300$\times$256 voxels.
Results of the analysis are presented in the form of  the same tables and figures as in Section \ref{Section3.2}.
 \label{table6}
\begin{table}
\caption{Model $M_{\rm ARs}$: volume metrics for $\bvec_{\rm calc}$ and
scalar potential $\phi$ computed for the entire volume $V$}
 \label{table6}
\begin{tabular}{cccccc}
\hline
  & $\left<\epsilon\right>_{\rm mean}$ & $\left<\epsilon\right>_{\rm median}$& $\epsilon_{\rm max}$& $\left<\epsilon\right>_{\rm w}$ & $\left<\epsilon\right>_{\phi}$\\
 \hline
Case I   &$3.758 \times 10^{-4}$&$2.573 \times 10^{-4}$&$4.955\times 10^{-1}$&$3.163\times 10^{-4}$&$1.219\times 10^{-3}$  \\
Case II   &$3.868\times 10^{-4}$&$2.473\times 10^{-4}$&$3.680\times 10^{-1}$&$2.170\times 10^{-4}$&$9.405\times 10^{-4}$  \\
Case III   &$2.154\times 10^{-2}$&$1.552\times 10^{-2}$&$1.809\times 10^{0}$&$6.932\times 10^{-3}$&$6.209\times 10^{-2}$  \\
 \hline
\end{tabular}
\end{table}

\begin{table}
\caption{Model $M_{\rm AR}$: surface metrics for $\bvec_{\rm calc}$ and
the scalar potential $\phi$ computed for the boundary surface $S$.}
 \label{table7}
\begin{tabular}{cccccc}
\hline
  & $\left<\epsilon\right>_{\rm mean}$ & $\left<\epsilon\right>_{\rm median}$& $\epsilon_{\rm max}$& $\left<\epsilon\right>_{\rm w}$ & $\left<\epsilon\right>_{\phi}$\\
\hline
Case I   &$2.475\times 10^{-2}$&$1.404\times 10^{-3}$&$8.130\times 10^{0}$&$4.523\times 10^{-2}$&$2.666\times 10^{-3}$  \\
Case II   &$5.935\times 10^{-2}$&$1.434\times 10^{-3}$&$1.358\times 10^{0}$&$4.002\times 10^{-3}$&$2.115\times 10^{-3}$  \\
Case III   &$7.682\times 10^{-2}$&$4.792\times 10^{-2}$&$1.308\times 10^{1}$&$1.078\times 10^{-1}$&$1.002\times 10^{-1}$  \\
 \hline
\end{tabular}
\end{table}

\begin{table}
\caption{Model $M_{\rm AR}$: volume metrics for $\curl\avec_{\rm calc}$ and the scalar potential $\phi$ computed for the entire volume $V$.}
 \label{table8}
\begin{tabular}{ccccc}
\hline
  & $\left<\epsilon\right>_{\rm mean}$ & $\left<\epsilon\right>_{\rm median}$& $\epsilon_{\rm max}$& $\left<\epsilon\right>_{\rm w}$ \\
 \hline
Case I   &$5.776\times 10^{-4}$&$2.816\times 10^{-4}$&$1.452\times 10^{0}$&$1.482\times 10^{-3}$  \\
Case II   &$6.602\times 10^{-4}$&$2.716\times 10^{-4}$&$2.052\times 10^{0}$&$1.643\times 10^{-3}$  \\
Case III   &$2.152\times 10^{-2}$&$1.550\times 10^{-2}$&$1.494\times 10^{0}$&$6.984\times 10^{-3}$  \\
 \hline
\end{tabular}
\end{table}
\begin{table}
\caption{Model $M_{\rm AR}$: surface metrics for $\curl\avec_{\rm calc}$ and the scalar potential $\phi$ computed for the boundary surface $S$.}
 \label{table9}
\begin{tabular}{ccccc}
\hline
  & $\left<\epsilon\right>_{\rm mean}$ & $\left<\epsilon\right>_{\rm median}$& $\epsilon_{\rm max}$& $\left<\epsilon\right>_{\rm w}$ \\
 \hline
Case I   &$4.304\times 10^{-2}$&$6.913\times 10^{-4}$&$1.326\times 10^{1}$&$9.649\times 10^{-2}$  \\
Case II   &$3.555\times 10^{-2}$&$8.725\times 10^{-4}$&$1.017\times 10^{1}$&$7.979\times 10^{-2}$  \\
Case III   &$8.471\times 10^{-2}$&$4.906\times 10^{-2}$&$1.549\times 10^{1}$&$1.276\times 10^{-1}$  \\
 \hline
\end{tabular}
\end{table}
\begin{table}
\caption{Model $M_{\rm AR}$: magnetic energy calculation results.}
 \label{table10}
\begin{tabular}{cccc}
\hline
  & $E$ & $E_\phi$& $E_{\rm virial}$ \\
 \hline
Model   &$7.7832\times 10^{32}$&$7.7766\times 10^{32}$&$7.7767\times 10^{32}$  \\
Case I   &$7.7828\times 10^{32}$&$7.7763\times 10^{32}$&$7.7764\times 10^{32}$  \\
Case II   &$7.7831\times 10^{32}$&$7.7765\times 10^{32}$&$7.7766\times 10^{32}$  \\
Case III   &$7.7718\times 10^{32}$&$7.7692\times 10^{32}$&$7.7262\times 10^{32}$  \\
 \hline
\end{tabular}
\end{table}
\begin{figure}
\centerline{\includegraphics[width=1.\textwidth,clip=]{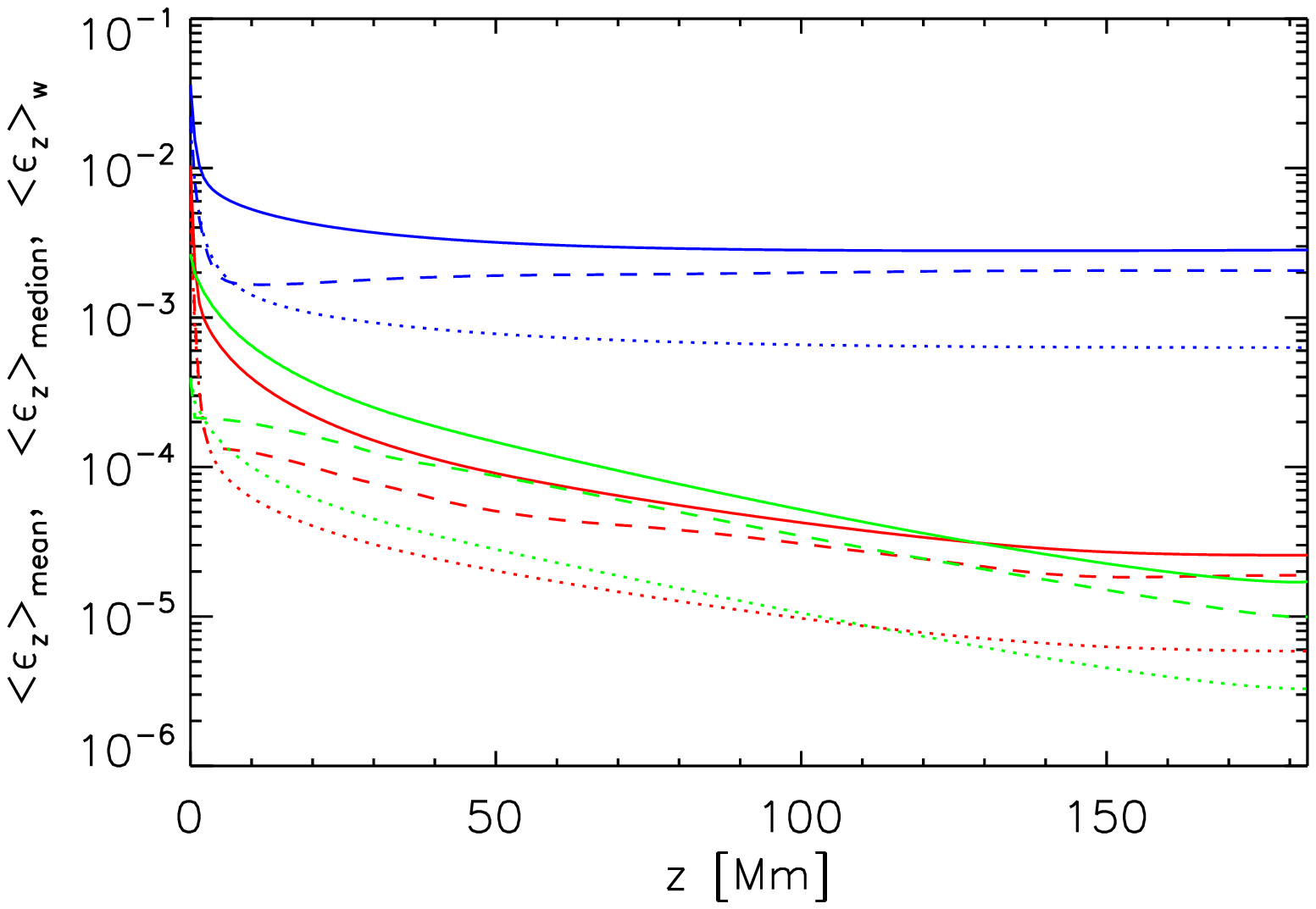}}
\caption{Dependence of the  average (solid line), median (dashed line) and the field weighted (dotted line) relative error  with respect to the model field ($\bvec_{\rm model}$) on the height ($z$) for the  potential magnetic field of the $M_{\rm AR}$ model, calculated with the Case I (red), Case II (green) and Case III (blue) methods from the Neumann boundary conditions.} \label{fig10}
\end{figure}
\begin{figure}
\centerline{\includegraphics[width=1.\textwidth,clip=]{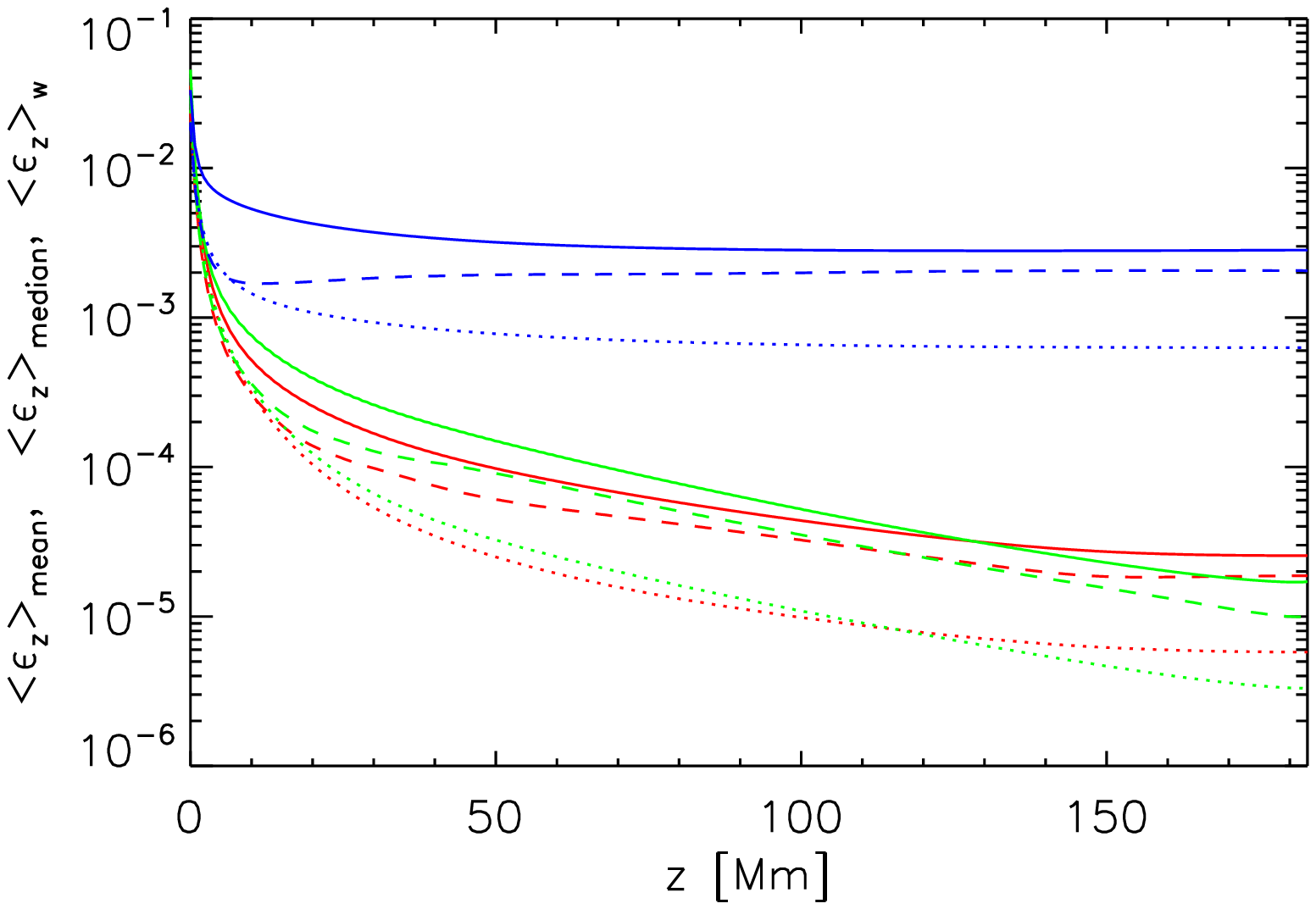}}
\caption{Dependence of the  average (solid line), median (dashed line) and the field weighted (dotted line) relative error with respect to the model field ($\bvec_{\rm model}$)  on the height ($z$) for the  potential magnetic field of the $M_{\rm AR}$ model,  recovered from the vector potential (${\bf \nabla}\times\avec_{\rm calc}$) calculated with the Case I (red), Case II (green) and Case III (blue) methods from the Neumann boundary conditions.} \label{fig11}
\end{figure}
\begin{figure}
\centerline{\includegraphics[width=1.\textwidth,clip=]{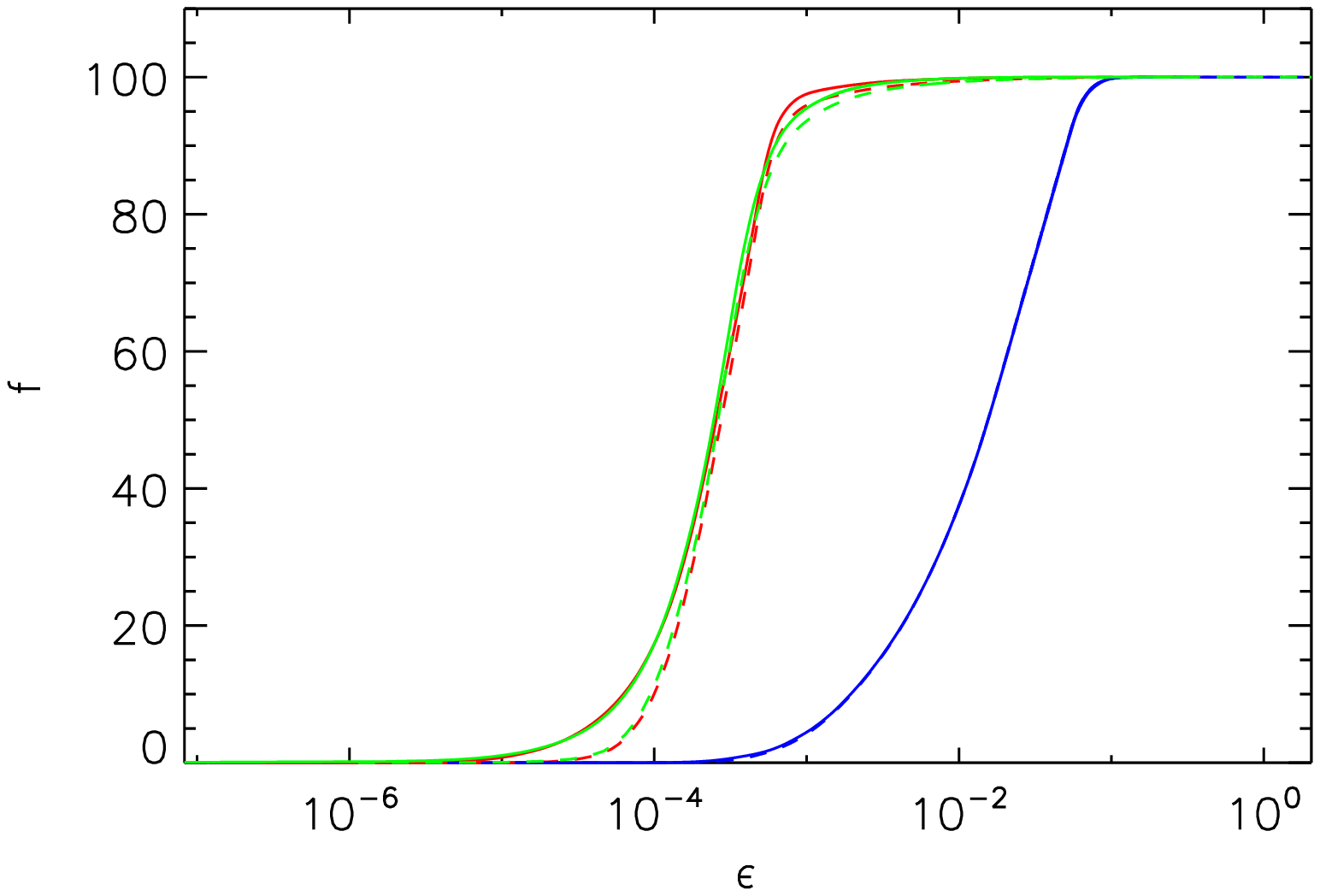}}
\caption{
    The distribution of the relative errors of the field $\bvec_{\rm calc}$ (solid line) and the field computed from the vector potential ${\bf \nabla}\times\avec_{c\rm alc}$  (dashed line) with respect to the model field $\bvec_{\rm model}$ for the potential field of the $M_{\rm AR}$ model calculated with the Case I (red), Case II (green) and Case III (blue) methods from the Neumann boundary conditions.
}
\label{fig12}
\end{figure}

Case I and Case II demonstrate similar accuracy for most of the metrics calculated for the $M_{\rm AR}$ model while the metrics for Case III are approximately two orders of magnitude worse.
The magnetic energy calculations are also much more accurate for Case I and Case II in comparison with the Case III results.

\subsection{Computing times for the three algorithms}
\label{Section3.4} All calculations are performed on a computer with an Intel(R) Core(TM) i5-4460 3.20GHz  processor.
 The computation times for different algorithms are the following:

 \begin{itemize}
\item For $M_{\rm charges}$
\begin{itemize}
    \item Case I -16.48 sec (without ${\bf \nabla} \times \avec $), 21.18 sec (with ${\bf \nabla} \times \avec $);
    \item Case II - 0.23 sec (without ${\bf \nabla} \times \avec $), 1.24 sec (with ${\bf \nabla} \times \avec $);
    \item Case III- 0.065 sec (without ${\bf \nabla} \times \avec $), 1.02 sec (with ${\bf \nabla} \times \avec $).
\end{itemize}
\item  For $M_{\rm AR}$
 \begin{itemize}
    \item Case I -13.6 h  (without ${\bf \nabla} \times \avec $), 27.4 h  (with ${\bf \nabla} \times \avec $); \\
    \item Case II - 202.91 sec (without ${\bf \nabla} \times \avec $), 424.25  sec (with ${\bf \nabla} \times \avec $); \\
    \item Case III - 0.93  sec (without ${\bf \nabla} \times \avec $), 253.05 sec (with ${\bf \nabla} \times \avec $). \\
\end{itemize}
\end{itemize}
\section{Conclusions}

We consider three different algorithms  for the numerical solution of the Neumann BVP for the Laplace equation in a 3D rectangular  box.
The first two options are based on  dividing the whole problem into six sub-problems corresponding to six faces of the box.
The solution of each sub-problem uses the ordinary Fourier decomposition of the solution into a set of harmonic functions satisfying the Laplace equation. In one variant (Case I) the solution is obtained through direct   integration  while in the other case (Case II) the FFT is applied to speed up the calculation.
The third algorithm (Case III) uses the proprietary Poisson solver from the Intel MKL library.
The solutions are presented as a set of three quantities, namely the magnetic field itself, and  its scalar and vector potentials. In Cases I and II all of the quantities are expressed and calculated as a superposition of the harmonic functions.  The Intel MKL library used in Case III allows for computation of the scalar potential only. The magnetic field vector is then computed using the finite difference scheme. The vector potential is  calculated from the field using the method proposed by  \cite{citeMVal}.

All three methods provide high accuracy of the potential field calculation. However, the Case I and Case II methods proposed here provide much more precise solutions with metrics which are around two orders of magnitude better than the Poisson solver supplied with the Intel MKL library. The  Case II method based on FFT provides the possibility of fast and accurate calculation of the potential field energy that is required for the estimation of the free magnetic in energy in solar active regions.  Although the direct spline based integration (Case II gives the best results among all considered methods, it is a very computationally expensive approach especially for high resolution grids. Hover, this algorithm can be used for  small and moderate-size computational domains. It also allows for potential field computation at arbitrary location, providing a continuous and precise solution. Although the Poison solver from the Intel MKL library (Case III) is not as precise as other approaches,  it is very fast and gives the result in less than a second even for the high resolution grids.

In conclusion, we  should note the quality of the algorithm for
the vector potential calculation \cite{citeMVal} that is used in
the Case~III method to compute the vector potential of a given
potential field. Our tests show that the reverse reconstruction of
the potential field from the vector potential does not lead to
significant worsening of the solution. It confirms the efficiency
of the method, that also can be applied in the more general case
of an arbitrary divergence-free field.

\begin{acks}
 This study was supported by the Russian Foundation of Basic Research under grants 15-02-01077, 16-32-00315\_mol\_a, 15-02-03835\_a, and 15-02-01089\_a; by the ISSI International Team on Magnetic Helicity estimations in models and observations of the solar magnetic field. This study
was supported by the Program of basic research of the RAS
Presidium No. 7.
\end{acks}

\begin{acks}[Disclosure of Potential Conflicts of Interest]
 The authors declare that they have no conflicts of interest.
\end{acks}


\end{article}

\end{document}